\documentclass[twocolumn,showpacs,prb,amsmath,amstex,amssymb,citeautoscript,longbibliography]{revtex4-1}
\pdfoutput=1
\usepackage{natbib}
\usepackage[english]{babel}
\usepackage{letltxmacro}
\LetLtxMacro{\ORIGselectlanguage}{\selectlanguage}
\makeatletter
\DeclareRobustCommand{\selectlanguage}[1]{%
  \@ifundefined{alias@\string#1}
    {\ORIGselectlanguage{#1}}
    {\begingroup\edef\x{\endgroup
       \noexpand\ORIGselectlanguage{\@nameuse{alias@#1}}}\x}%
}
\newcommand{\definelanguagealias}[2]{%
  \@namedef{alias@#1}{#2}%
}
\makeatother
\definelanguagealias{en}{english}
\definelanguagealias{English}{english}

\usepackage{float}
\usepackage{graphicx}
\usepackage[usenames,dvipsnames]{color}

\usepackage[colorlinks=true,allcolors=blue]{hyperref}

\usepackage{bbm}

\newcommand{\gs}{|\emptyset\rangle}
\newcommand{\kk}{\Delta}

\newcommand{\tr}[1]{\,\mathrm{tr}[#1]}	
\newcommand{\mean}[1]{\langle #1 \rangle}	

\newcommand{\abs}[1]{\ensuremath{\left\vert #1 \right \vert}} 
\newcommand{\bra}[1]{\langle #1 |}
\newcommand{\ket}[1]{| #1 \rangle}
\newcommand{\braket}[2]{\langle #1|#2 \rangle}

\newcommand{\tot}[1]{\text{d}\hspace{-0.8pt}#1\;}

\newcommand{\tone}{\mathbbm{1}}

\def\clap#1{\hbox to 0pt{\hss#1\hss}}

\def\mathclap{\mathpalette\mathclapinternal}

\def\mathclapinternal#1#2{%
\clap{$\mathsurround=0pt#1{#2}$}}

\newcommand{\refapp}[1]{App.~\ref{#1}}

\begin{document}

\title{Non-interacting central site model: localization and logarithmic entanglement growth}

\author{Daniel Hetterich$^1$}
\author{Maksym Serbyn$^2$}
\author{Fernando Dom\'inguez$^1$}
\author{Frank Pollmann$^3$}
\author{Bj\"orn Trauzettel$^1$}
\affiliation{${}^1$Institut f\"{u}r Theoretische Physik, Universit\"{a}t W\"urzburg, D-97074 W\"urzburg, Germany}
\affiliation{${}^2$Department of Physics, University of California, Berkeley, California 94720, USA\\ and Institute of Science and Technology (IST) Austria, 3400 Klosterneuburg, Austria}
\affiliation{${}^3$Max-Plank Institute for the Physics of Complex Systems, D-0118 Dresden, Germany}
\date{\today}

\begin{abstract}
We investigate the stationary and dynamical behavior of an Anderson localized chain coupled to a single central bound state. Although this coupling partially dilutes the Anderson localized peaks towards nearly resonant sites, the most weight of the original peaks remain unchanged. This leads to multifractal wavefunctions with a frozen spectrum of fractal dimensions, which is characteristic for localized phases in models with power-law hopping. Using a perturbative approach we identify two different dynamical regimes. At weak couplings to the central site, the transport of particles and information is logarithmic in time, a feature usually attributed to many-body localization. We connect such transport to the persistence of the Poisson statistics of level spacings in parts of the spectrum. In contrast, at stronger couplings the level repulsion is established in the entire spectrum, the problem can be mapped to the Fano resonance, and the transport is ballistic.
\end{abstract}

\maketitle

\section{Introduction}
In quantum mechanics the destructive interference of wave
functions in the presence of disorder may completely suppress
diffusion, leading to the celebrated Anderson localization
(AL)~\cite{Anderson1958}.  AL is driven by a competition
between the disorder potential and the kinetic energy, and strongly depends on the spatial
dimension~\cite{Kramer1993}. The Anderson transition
(AT) between metallic and localized phases in disordered
systems presents an intriguing example of ergodicity
breaking in non-interacting system, and has been subject
of theoretical~\cite{Evers2008} and experimental studies \cite{Chabe2008} for many years.

The search for delocalization transitions in one-dimensional
systems with uncorrelated disorder motivated the introduction
of models with power-law hoppings. In particular, the
power-law banded matrices~\cite{Mirlin2000}
became a prime example allowing for a detailed study of AT
criticality.  Further, random-matrix-type models without any
spatial structure were recently
considered~\cite{Kravtsov2015}.
The power-law banded matrices can be viewed as a
one-dimensional system with a  power-law hopping,
$1/r^\sigma$, controlled by the exponent $\sigma$.
Such systems are localized when $\sigma>1$, delocalized
for $\sigma<1$, and right at the AT for $\sigma=1$.

In this work, we study spectral and transport properties of the central site model (CSM), in which a single site is coupled to each site of an AL chain underlying random potentials (see inset in Fig.~\ref{fig:loclength}). 
The CSM effectively combines long range hoppings and disorder, which are known to be key features of power-law banded matrices. Hence, it is an ideal candidate for studying the physics of AL. 
This model is further motivated by central spin models,\cite{Loss1998a,Schliemann2002,Bortz2007} which adequately describe the hyperfine interaction between an electron spin localized in a quantum dot with the bath of nuclear spins in the host material~\cite{Coish2004, Fischer2009}.
Related to central spin models, the Kondo effect has been analyzed in the presence of disordered metals. 
In particular, the authors of Ref.~\onlinecite{Kettemann2007} found that the probability for a magnetic moment to remain free down to zero temperature increases with disorder strength.

The CSM formally resembles the Fano resonance problem~\cite{Mahan1990a}. 
However, the presence of disorder within the continuum leads to novel physics. 
We find multifractal wave functions with a frozen spectrum for any finite coupling to the central site.
This property is characteristic to AL in systems with power-law hopping~\cite{Mirlin2000}. 
In our model, the coupling of the AL chain to the central site allows to explore critical particle transport, which is intimately connected to the spread of entanglement.

We identify two different regimes depending on the coupling strength. 
At weak coupling, the CSM retains the Poisson level statistics
coming from the AL chain in most parts of the spectrum. We demonstrate that the absence of level repulsion 
leads to a logarithmic in time growth of entanglement entropy.
Remarkably, this feature is usually attributed to the interacting many-body localized phase~\cite{Znidaric2008,Bardarson2012,Serbyn2013}.
At strong coupling instead, level repulsion is recovered in the whole spectrum,
leading to similar physics as in the Fano resonance problem, where the entanglement entropy grows linearly in time.

The article is organized as follows. In Sec.~\ref{sec:model}, we introduce the specific model studied in this paper. Next,  in Sec.~\ref{sec:eigenfunctions},  we study the structure of the eigenfunctions under the influence of the central site. There, beside analyzing how much weight of the original Anderson peak spreads through the system, we also characterize the eigenfunctions by means of multifractal properties. The observed statistics of eigenvalues is, as we present in Sec.~\ref{sec:level_statistics}, a competion between the Poisson-distributed energies of the localized chain and the level splitting induced by the central site. In Sec.~\ref{sec:entanglement_growth}, we describe the dynamical properties of particles in the CSM and the related entanglement entropy.
For a better readability, we delegate the details of the calculations to the  appendices.

\section{The central site model} 
\label{sec:model}
The CSM is described by the quadratic Hamiltonian $H=H_\text{ring}+H_\text{c}$ that consists of two parts.
The first one describes a one-dimensional disordered ring of
size $L$, 
\begin{equation}\label{Eq:Hring}
H_{\rm{ring}}=\sum_{i=1}^L h_{i} c_i^\dagger c_i + J\left(c_i^\dagger c_{i+1} + c_{i+1}^\dagger c_i\right),
\end{equation}
 where $c_i^\dagger$ and $c_i$ are fermionic creation and annihilation operators.
In $H_\text{ring}$, the on-site ring energies $h_i$ are uniformly distributed random values $h_i\in \left[-W, W\right]$, where $W$ quantifies
the disorder strength and is assumed to be dominant over the
hopping $J$, $W>J$.

\begin{figure}
\centering
\includegraphics[width = 0.98\linewidth]{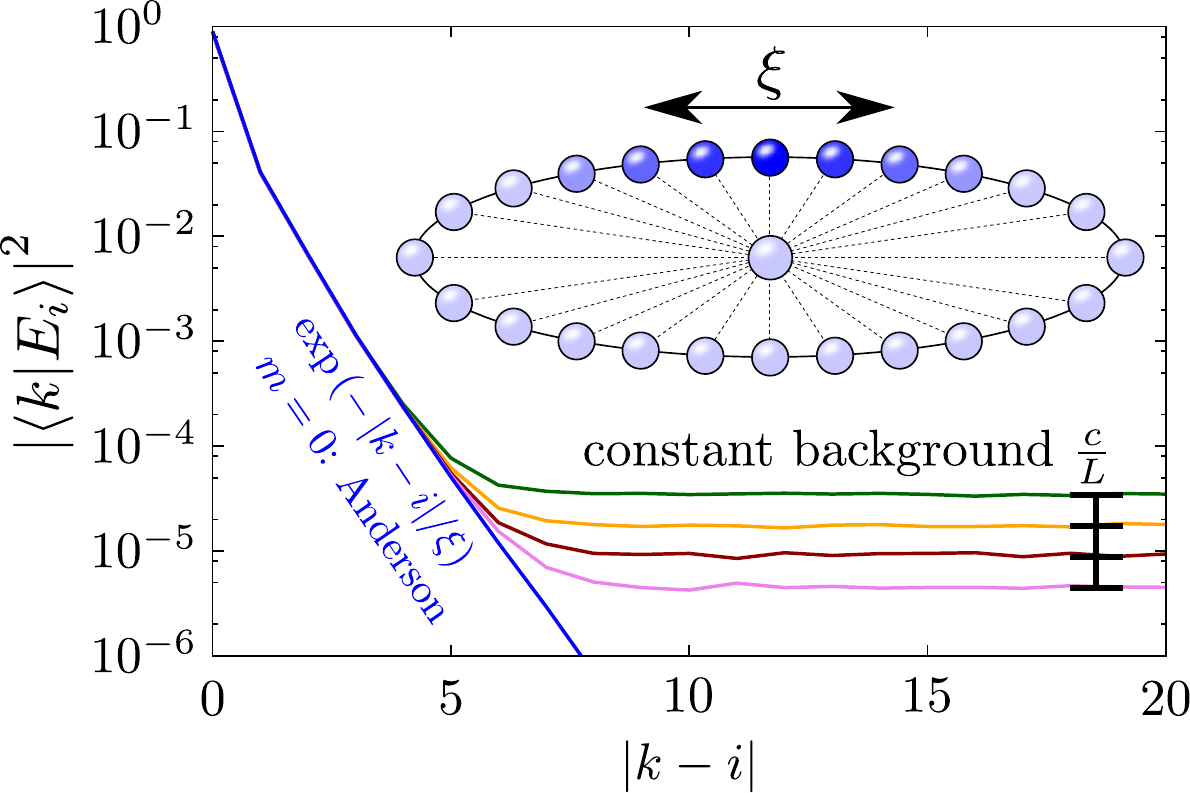}
\caption{
The wave function of the eigenstate that is localized
on the site $i$ initially decays exponentially with
the distance $\abs{k-i}$ with the same localization
length as in the model without central site (blue line).
Non-zero hopping to the central site, $m>0$ leads to a
saturation of the wave function amplitude to a constant
background that scales as~$c/L$. We show data for disorder strength $W=10$, coupling strength $m=1.25$ and $L\in\{2^9,2^{10},2^{11},2^{12}\}$.
The inset displays the schematic of the CSM, dashed lines represent the hopping terms $m/\sqrt{L}$.
}
\label{fig:loclength}
\end{figure}

The second part is referring to the Hamiltonian of the
central site and its coupling to all ring sites,
\begin{equation}\label{Eq:Hc}
H_\text{c} =h_0 c_0^\dagger c_0+ \frac{m}{\sqrt{L}}\sum_{i=1}^L \left( c_i^\dagger c_0 + c_0^\dagger c_i\right),
\end{equation}
where in what follows we set the energy of the central
site $h_0=0$. 
Due to this star-like coupling (see inset of Fig.~\ref{fig:loclength}) all sites are at most next-nearest neighbors such that there is no concept of distance.  
Note, that we scale the coupling
to the central site as $1/\sqrt{L}$, so that our system
has a well defined thermodynamic limit.
In the remainder of the paper we set $J=\hbar=1$, and
measure energy and time in units of $J$ and $\hbar/J$, respectively.

\section{Multifractal structure of eigenfunctions}
\label{sec:eigenfunctions}
\label{sec:multifractality}
We study the eigenfuctions $\ket{E_i}$, which are localized at site $i$, in real space by using numerical exact diagonalization techniques. 
We are especially interested in how the original Anderson localization peaks are modified by the central site. 
For this purpose, we discuss the average probability $\abs{\braket{k}{E_i}}^2$ of measuring a particle in the eigenstate $\ket{E_i}$ at site $k$ in
 Fig.~\ref{fig:loclength}.
We find that the eigenfunctions are still exponentially localized around the $i$th site, with the length scale $\xi$.
This localization length is independent of the coupling $m$ and coincides with the known AL length of $H_{\rm{ring}}$.\cite{Anderson1958, Giamarchi1988, Kramer1993}
Thus, the main effect on the structure of the eigenfunctions of the coupling to the central site is the small, on average homogenous, background shown in Fig.~\ref{fig:loclength}.

Since this is a non-interacting problem we can solve for the wave functions self-consistently. The mixing of localized wave functions of the decoupled AL ring goes through the central site which has a constant spectral weight dissolved in the continuum irrespective of the coupling $m$ (see \refapp{app:fano}). This peculiarity of the CSM leads to a constant background of the wave functions displaced away from the initial site. Thus, the background has to scale as $c/L$, which coincides with our numerics, see  Fig.~\ref{fig:loclength}. 
In \refapp{app:three_site}, we derive that  $c\sim m^2/W^2$ for $m\ll W$.
Note that due to this scaling of the constant background, the constant $c$ can be interpreted as the probability, that a particle in the state $\ket{E_i}$ can be found outside the Anderson peak, which is independent of the system size $L$.

\begin{figure}[t]
\centering
\includegraphics[width = 0.95\linewidth]{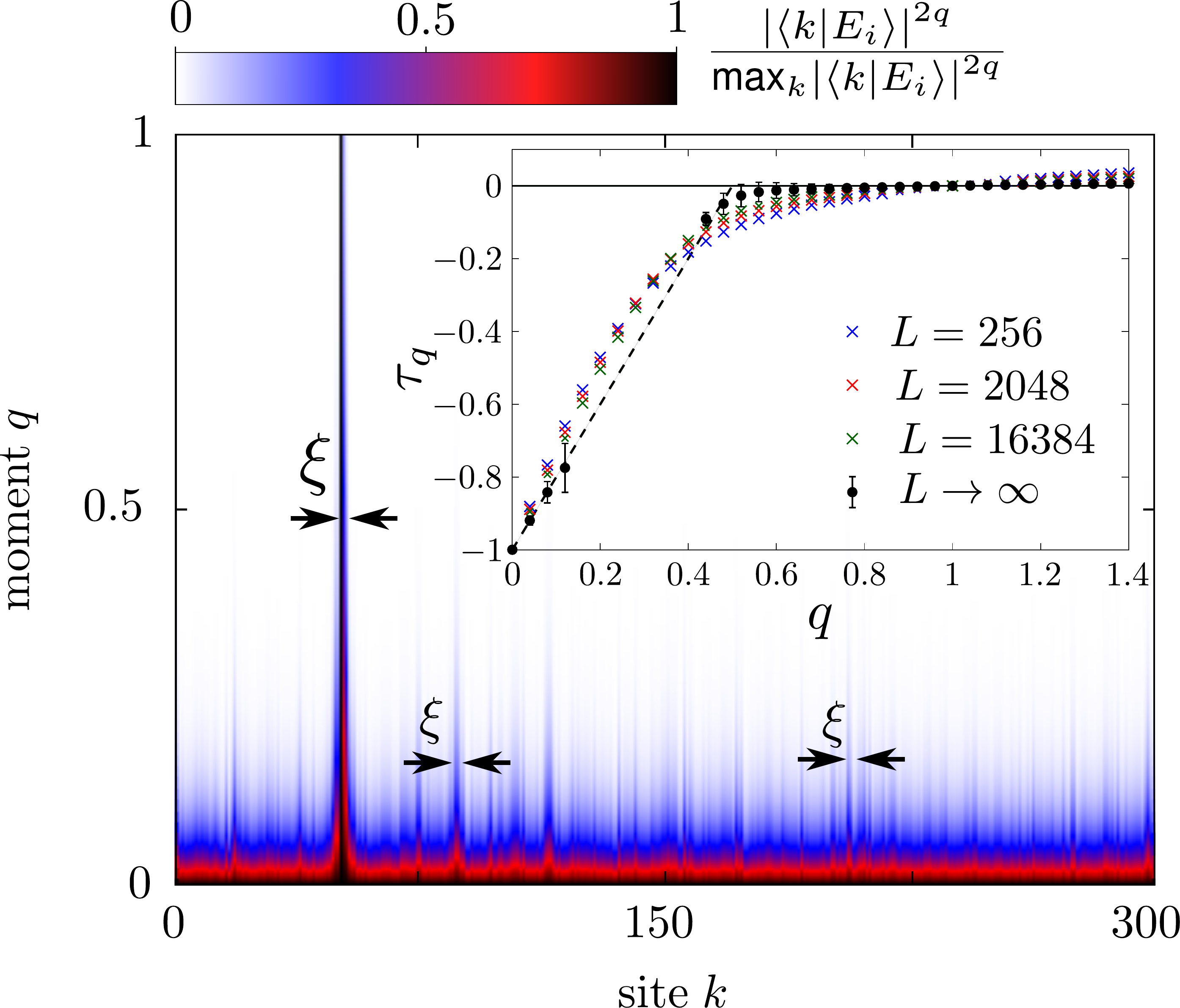}
\caption{
Multifractality of wave functions in real space. We show continuous moments $q\in [0,1]$ of the distribution of an arbitrary eigenstate $\ket{E_i}$ in the middle of the spectrum. Besides the original localization center (around $k\approx 50$), wave functions peak at resonantly coupled sites if $m>0$, which establishes fractal structures of Anderson peaks.
The inset shows the spectrum of fractal dimension $\tau_q$, which converges to $\tau_q=0$ when $q \geq 1$ (spectral freezing \cite{Mirlin2000}). Data is for eigenstates of the energy interval $\mathcal{I}=[2,4]$ at disorder strength $W=10$ coupling $m=1$ and system sizes with up to $L=23170$. Black dots show $\tau_q$  extrapolated to the thermodynamic limit. The error bars reflect the uncertainty from determining parameters of the fitting function. Due to the non-monotinicty of $\tau_q$ with $L$ in the interval $q \in [0.15, 0.4]$ the scaling analysis does not work. However, from the convexity and monotonicity of $\tau_q$, we expect the dashed line, which is $1 + 2q$, to describe the statistics.
For all values of $m$ we receive qualitatively similar fractal statistics after the scaling.}
\label{fig:multifractality}
\end{figure}

The constant background in Fig.~\ref{fig:loclength} results from the averaging of individual eigenstates $\ket{E_i}$ over different disorder configurations. For each separate $\ket{E_i}$, the coupling to the central site creates smaller duplicates of the original Anderson peak at all resonant sites, {\it cf.}  Fig.~\ref{fig:multifractality}, which leads to multifractal structures \cite{Schreiber1991}. These are characterized by means of the moments of the participation ratios,
\begin{align}\label{Eq:P-def}
P_q = \mean{\sum_{k} \abs{ \braket{k}{E_i}}^{2q}}_i \sim L^{-\tau_q},
\end{align}
which we average over all eigenstates $\ket{E_i}$ and over disorder realizations.
The scaling $P_q \sim L^{-\tau_q}$ defines $\tau_q= D_q(q-1)$ and the fractal dimension $D_q$ \cite{Schreiber1991}.
For the case of ideal metals or insulators, one expects the constant values $D_q = d$ and $D_q=0$, respectively, where $d$ is the spatial dimension.  Instead,  multifractal wave functions exhibit a  dependency on $q$.

The inset of Fig.~\ref{fig:multifractality} demonstrates that the coupling to the central site partially destroys the insulating phase and gives rise to a non-trivial spectrum $\tau_q$. Nevertheless, for all values of $m$ we find $\tau_q=0$ for $q\geq1$.  This convergence refers to a  ``frozen'' fractal spectrum.
This observation is consistent with the presence of the AL peak in Fig.~\ref{fig:loclength}, because for $q\gtrsim 1$, all participation ratios are given by the  largest values of the wave function.

Although the original Anderson peak dominates the fractal statistics at $q\gtrsim 1$, its fractal replicas at resonant sites (see Fig.~\ref{fig:multifractality}) become more important at smaller values of $q$ (a pure Anderson insulator shows $\tau_q=0$ for all $q>0$). Below we will use the fact that the physics of the CSM is well approximated by studying the interplay between these resonant sites. Furthermore, we show that the number and size of the Anderson peak replicas can be altered by modifying the strength of the coupling to the the central site $m$. This allows for two different regimes where transport properties and eigenvalue statistics either resemble metals or are or critical (see below).

\section{Level statistics}
\label{sec:level_statistics}
\begin{figure}
\centering
\includegraphics[width = 0.99\linewidth]{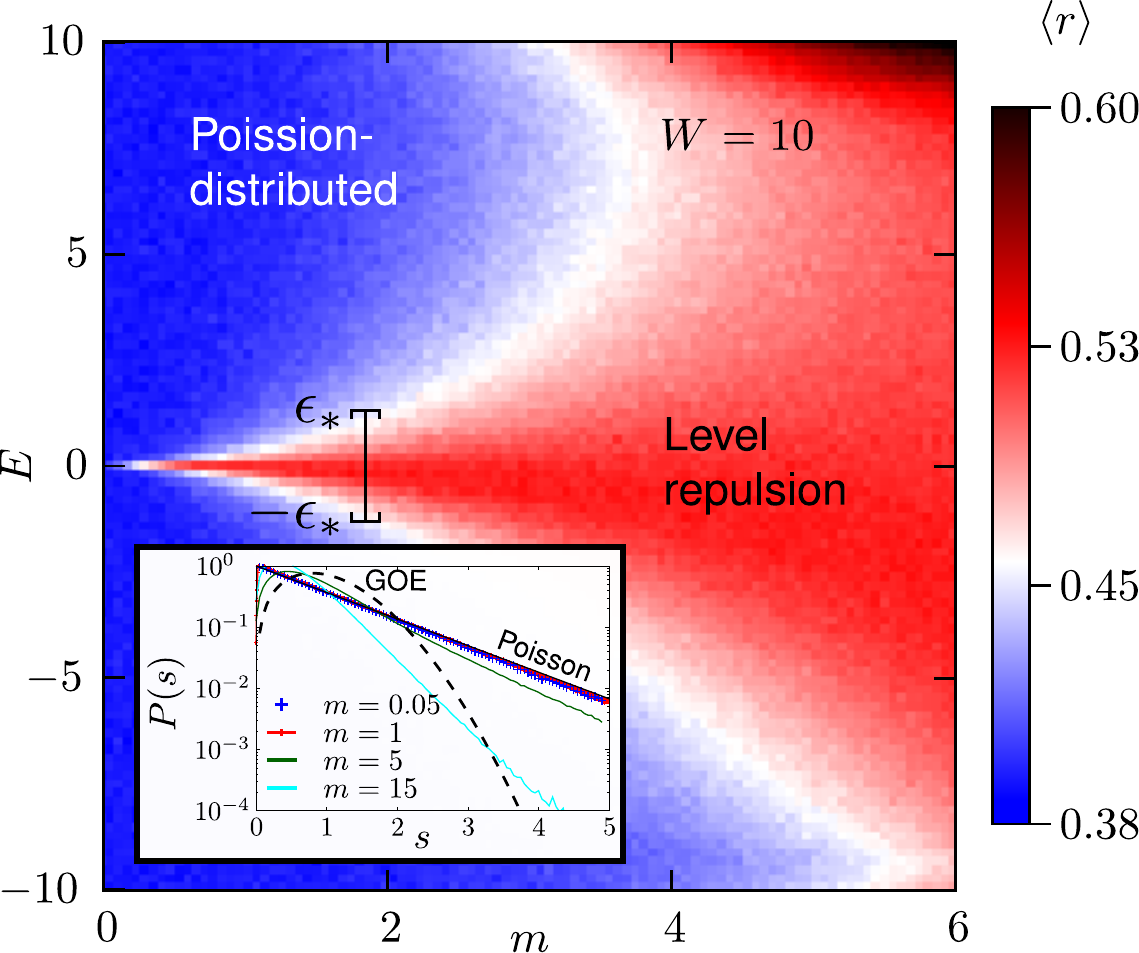}
\caption{
Energy resolved distribution of eigenvalues $E_i$ for different coupling strengths $m$. The color quantifies the disorder averaged value of $r = \min{(g_i,g_{i+1})}/\max{(g_i,g_{i+1})}$, where $g_i = E_{i+1}-E_i$ is the gap between adjacent eigenvalues at energy $E$ for fixed $m$. The interval $[-\epsilon_*,\epsilon_*]$, in which the eigenvalues repel each other, growths with increasing $m$. The data, generated at $L=2^{11}$ and $W=10$, ranges from $\mean{r}=0.38$ (Poisson-distribution) to values that exceed $0.53$, which would be typical for the GOE. In the inset we demonstrate that the CSM does indeed not provide GOE distributions. Even in regimes with strong level repulsion, the tails of the  distribution  $P(s)$ of gaps $s_i=g_i/\delta$ decrease exponentially rather than Gaussian (dashed line). 
}
\label{fig:ps}
\end{figure}

While we have focused on the shape of the eigenstates in the last section, we  analyze now how the corresponding eigenvalues are distributed as a function of the coupling to the central site.
In the absence of the central site, $m=0$, the eigenvalues of the AL chain are Poisson distributed.
For finite values of $m$, eigenvalues that are energetically close to the potential of the central site ($h_0 =0$) repel each other. 
Specifically, an eigenstate of the AL chain with energy $\epsilon$ will acquire a correction  $m^2/(\epsilon L)$ within second order perturbation theory. 
Then, we define the crossover energy $\epsilon_* =  m^2 / \delta L$, where the energy correction is of similar size as the mean level spacing $\delta$.
Hence, only the eigenvalues within the interval $[-\epsilon_*, \epsilon_*]$ acquire level repulsion, while the eigenvalues outside of this interval remain Poisson distributed.
\vspace{1cm}\\

In Fig.~\ref{fig:ps} we use an energy resolved analysis of eigenvalue statistics to illustrate how the region of level repulsion increases with $m$.
At weak coupling $m\ll W$, the amount of levels repelling each other, shown in red color, is negligible and Poisson statistics dominates the spectrum (blue areas). Although the spectrum resembles the AL distribution, the presence of any finite region of level repulsion induces critical transport properties in the CSM, which we study below. 
For $m \gtrsim W$, $[-\epsilon_*, \epsilon_*]$ covers the whole spectrum such that all previously resonant levels are split due to level repulsion. We show in \refapp{app:fano} that, in this limit, the conventional Fano resonance picture~\cite{Mahan1990a} is capable of describing the CSM.

The coupling to the central site establishes level repulsion, such that the probability $P(s)$ of finding a gap $s_i=(E_i-E_{i+1})/\delta$ between to adjacent eigenvalues vanishes for $s\to 0$.
However, the CSM cannot be brought into a region where the eigenvalues are Gaussian orthogonal (GOE) distributed, $P(s)= \frac{\pi}{2} s e^{-s^2\pi^2/4}$. Instead, $P(s)$ retains exponential tails of Poisson distributed systems, see inset of Fig.~\ref{fig:ps}. 
The exponential instead of Gaussian decay of $P(s)$ of large gaps $s$ suggests a finite level compressibility \cite{Altshuler1986, Chalker1996}. This is natural: while a single central site can easily split degeneracies causing the level repulsion at small $s$, it cannot lead to a strong mixing of eigenstates with very different energies.  Hence the fluctuations in the level spacing between $k$-th nearest neighbor eigenstates would grow proportionally to $k$, corresponding to finite compressibility.  In this sense, the level statistics is similar to the critical level statistics seen at the AT~\cite{Shklovskii1993, Kravtsov1994, Mirlin2000a}.

\section{Dynamics and entanglement growth}
\label{sec:entanglement_growth}

\begin{figure}
\centering
\includegraphics[width = 0.99\linewidth]{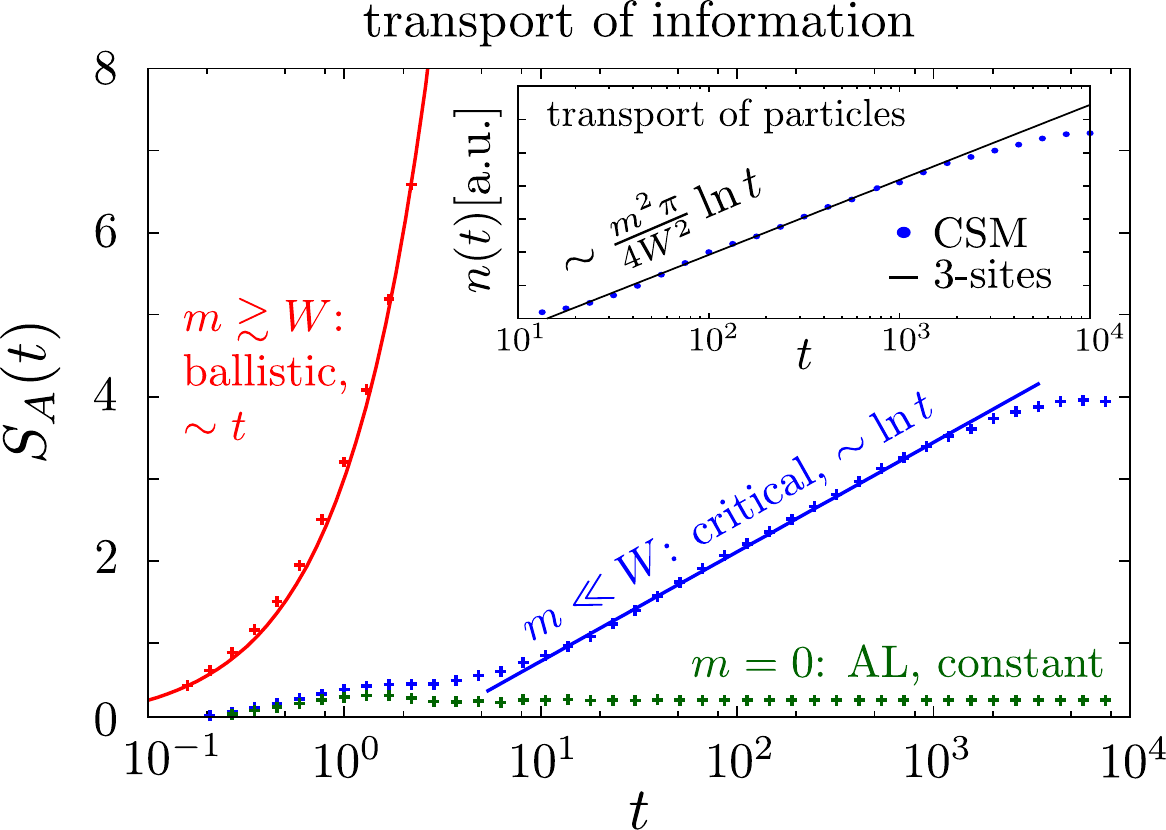}
\caption{Entanglement growth of the initial product state with $L/2$ fermions, where all even sites are occupied, $\ket{0101\ldots01}$. Upon changing the value of $m$, we observe Anderson insulating behavior ($m=0$, central site is decoupled), a regime of logarithmic entanglement growth ($W\gg m>0$), and eventually a  polynomial in time entanglement dynamics for $m\geq W$. The data is for $L=512$ ring sites, disorder strength $W=10$, and coupling $m=0$ (green), $m=1$ (blue), and $m=10$ (red).
The inset shows the logarithmic motion of the particles for $m=1$ and a comparison with the result of our analytical prediction.
}
\label{fig:entgrowth}
\end{figure}

In this section, we study the influence of the coupling to the central site $m$ to the dynamical properties of the model. 
To this end, we study the motion of particles and the spreading of information, witnessed by the entanglement growth.
Note that these two quantities are closely related in noninteracting models. In the CSM, particles are free, and hence, information from a subsystem $A$ can only be transferred to a disjunct subsystem $B$ if a particle moves in this direction.

In order to quantify the transport of information we consider the time evolution of the entanglement entropy $S_A(t)$ \cite{Calabrese2005, Luitz2014, Kaufman2016a}, which is frequently analyzed to identify the localization transition.
Using an equal size bipartition $A$, $B$ of the Hilbert space $ \mathcal{H}$ in real space, we quantify the correlations among them by studying the von Neumann entropy $S_A(t)$ of the reduced density matrix.
The evolution of $S_A(t)$ provides then a direct measure of the spread of information throughout the system.
In noninteracting systems,  $S_A(t)$ can be written in terms of the correlation matrix $C_{ij}^A = \bra{\psi(t)} c_i^\dagger c_j \ket{\psi(t)}$~\cite{Cheong2004,Peschel2002}, with $i,j\in A$, as
\begin{equation}
S_A(t)= -\tr{C^A \ln  C^A+(1-C^A)\ln(1-C^A)}.
\label{eq:sa}
\end{equation}
For the transport of particles, we place a single particle at an arbitrary ring site $i$ and calculate the amount 
\begin{equation}
n(t) = \mean{1- \abs{\bra{\psi(t)} c_i^\dagger c_i \ket{\psi(t)}}^2}_i
\end{equation}
of the particle that left the initial site. At $m=0$, the central site does not influence the AL ring such that transport of both, particles and information, over length scales that exceed the localization length $\xi$ is absent \cite{Anderson1958} . For finite $m$, the increasing region of level repulsion in the spectrum of eigenvalues, or likewise, the growing contribution of multifractal Anderson peak replicas that we studied in the previous sections, enables transport that can be characterized within two different regimes. These regimes will now be discussed seperately.

\begin{figure}
\centering
\includegraphics[width = 0.945\linewidth]{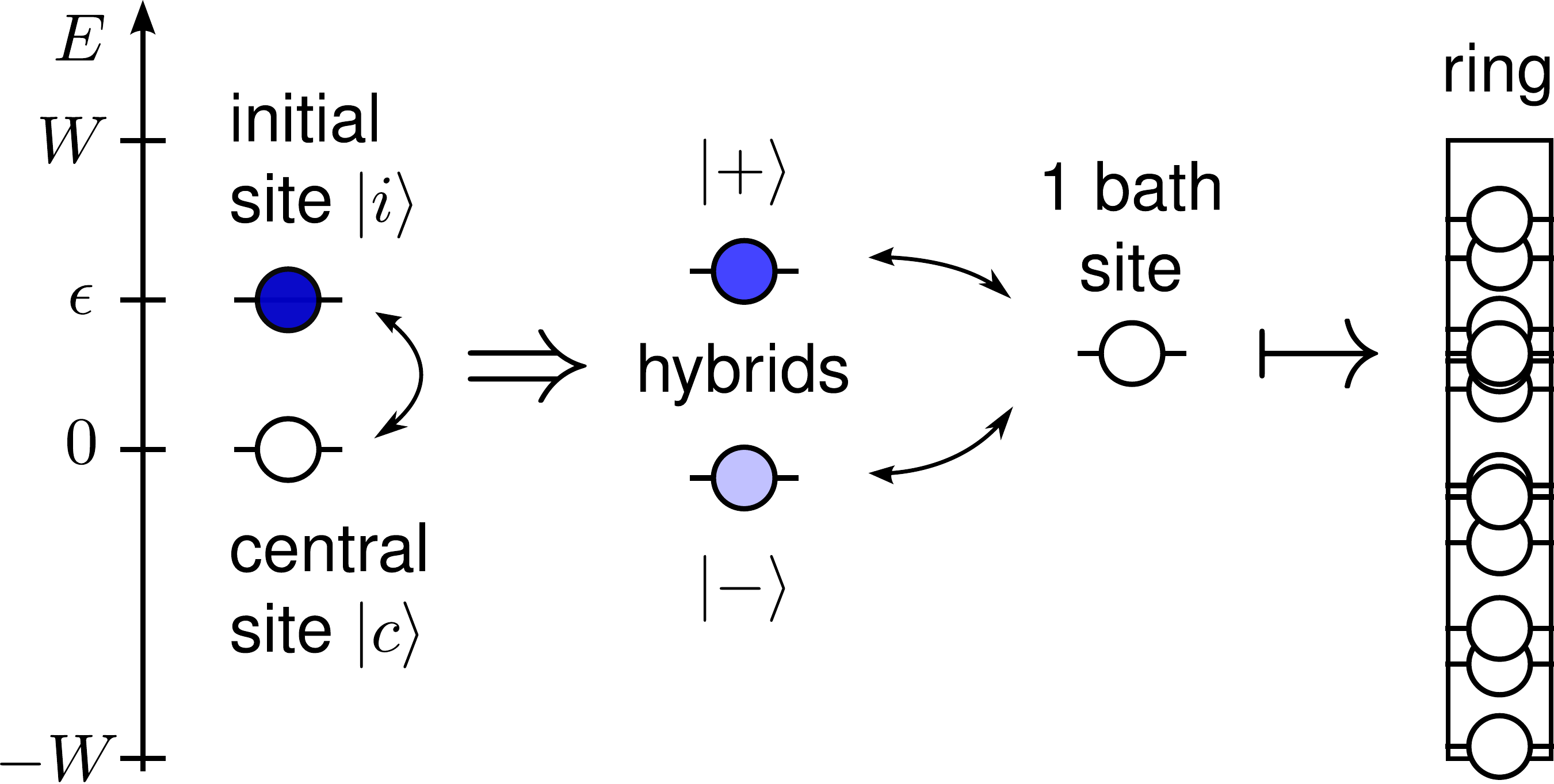}
\caption{Schematic of the effective three-site model. A free fermion is placed on an initial site $\ket{i}$ of potential $\epsilon$. This site couples with the central site $\ket{c}$ at potential $0$ and forms the two hybrids $\ket{\pm} \propto \ket{i} + a_{\pm}(m) \ket{c}$ shown in the figure. Both hybrids are then perturbatively coupled to a single final site of the remaining ring. After calculating the transport from the initial site to the final site, we average the potentials of these two sites over the energy interval $[-W,W]$ in order to mimic the whole ring of sites. With this toy model we obtain the quantitatively correct transport behavior of the CSM at low coupling constants $m\ll W$ as shown in the inset of Fig.~\ref{fig:entgrowth}.}
\label{fig:threesites}
\end{figure}

\subsection{Critical Transport, $m\ll W$}
In Fig.~\ref{fig:entgrowth}, we show  the time evolution of $S_A(t)$ in  
three regimes of the coupling strength $m$. 
In the case of non-zero but small $m\ll W$ (blue line), we
find a logarithmic dependence of $S_A(t)$ on time and a finite size saturation value that scales with $L$. These features have been associated so far to MBL phases where
information spreads via interactions between the particles~\cite{Bardarson2012,Serbyn2013}.
To the best of our knowledge, the logarithmic growth together with a saturation value that scales linearly with system size has never been observed in a non-MBL system before. Recently, a similar situation in the absence of interactions but without a linear scaling with system size has been identified in a different model  \cite{Singh2016}.

In order to understand the peculiar entanglement dynamics, we study how particles move through the central site that connects any possible bipartition.  Thus,
we consider how a particle placed at a specific site on the ring evolves with the CSM Hamiltonian.
Let us assume that this initial site has the strongest
overlap with the eigenstate of energy $\epsilon$ of the unperturbed chain.
The slowest dynamics is then generated by the mixing with almost degenerate energy
levels. The coupling between these neighbored levels is according to perturbation theory $J_\text{nn}\sim m^2/(\epsilon
L)$.
Hence, the particle leaks into such an eigenstate on the timescale $t
(\epsilon) \sim \epsilon L/m^2$.
Eigenstates with $\epsilon>\epsilon_*$ are hardly perturbed by the central site. However, in rare cases with probability $J_\text{nn} /\delta$, adjacent eigenstates are close to each other and strongly mixed by the central site. Then, the particle is equally likely to be found in both eigenstates at large times, contributing to a probability that the particle left its initial eigenstate given by $\bar{n}_\epsilon \sim J_\text{nn}/\delta \sim m^2/(W \epsilon)$.

The above intuition is formalized by a three-site model. 
Within this toy-model, we study the motion of a particle starting on a site
of potential $\epsilon$ through the central site into the continuum (see Fig.~\ref{fig:threesites} for the scheme). As the
coupling to the central site is weak, $m\ll W$, we are in the perturbative regime where mixing occurs with a small number of sites within $[-\epsilon_*,\epsilon_*]$.  Hence, in our case, it suffices to consider the coupling to a single site of energy $\epsilon'$ and average over all $\epsilon' \in [-W,W]$ after the equations of motions are solved. This average neglects correlations between initially unoccupied ring sites, which is a reasonable assumption for $m\ll W$. However, it accounts for an arbitrarily small level splitting which is  present in our problem.
 Using the three-site model, see \refapp{app:three_site}, the probability $n_\epsilon(t)$ that the particle left its initial site becomes
\begin{equation}
n_\epsilon(t) \approx \begin{cases}
\frac{\pi m^4}{W L }\left(\frac{t}{\epsilon^2}\right), & \frac{\sqrt{6}}{\epsilon}  \leq  t \leq \frac{\epsilon L}{4m^2}, \\
\bar{n}_\epsilon = \frac{\pi m^2}{4 W }\left(\frac{1}{\epsilon}\right), & t \geq \frac{L \epsilon}{4m^2},
\label{eq:numerical_evidence}
\end{cases}
\end{equation}
which depends on the potential at the initial position.
At a given time $t$, sites with energy $\epsilon< \epsilon_t= 4m^2 t/L$ have saturated their dynamics. Summing up the contributions from all such sites, we get that the number of particles that left their initial site is 
\begin{equation}
n(t)=\frac{1}{2W} \int_{-\epsilon_t}^{\epsilon_t} n_\epsilon(t) \tot{\epsilon}
= \frac{\pi m^2}{4W^2} \ln\left(\frac{m^2 }{L}t \right)
\label{eq:n3s}
\end{equation}
to leading order. In the inset of Fig.~\ref{fig:entgrowth}, we compare the result of the three-site model, Eq.~(\ref{eq:n3s}), with the numerical data of the full model and find  good agreement in the regime of logarithmic growth of entanglement. As we outline in \refapp{app:transport_to_entanglement}, the entanglement entropy $S_A(t)$ can be expressed by means of $n_\epsilon(t)$, which also explains  the logarithmic time-dependence of the entanglement entropy, plotted in Fig.~\ref{fig:entgrowth}.
The linear scaling of the saturation value of $S_A(t)$ and $n(t)$ for $t\to\infty $ with system size for all $m>0$ is derived in \refapp{app:fano} and  \refapp{app:three_site}.

\subsection{Linear Transport, $m\gtrsim W$}
Increasing the coupling constant to values $m\gtrsim W$, the CSM shows a linear time-dependence of $S_A(t)$ (see Fig.~\ref{fig:entgrowth}, red), which is typical for systems with level repulsion \cite{Lieb1972, Calabrese2005}.
One can readily understand this behavior with the previously defined quantities.
At $m\geq W$, the energy scale $\epsilon_*$ becomes of the order of the bandwidth $W$, such  level repulsion is present for all energy states of the Hamiltonian.
Then, we expect much faster spreading of particles, caused by the strong level admixture: the saturation time in such regime becomes simply an inverse level spacing $1/\delta$. Inserting this into the first line of  Eq.~(\ref{eq:numerical_evidence}), we obtain $\bar n_\epsilon\propto 1/\epsilon^2$. Averaging over the initial energy $\epsilon$ results then in the linear spreading of $n(t)$, and ballistic entanglement dynamics, as confirmed in Fig.~\ref{fig:entgrowth}. We further derive this linear growth of $n(t)$ analytically  in \refapp{app:fano} by means of a self-consistent perturbation theory within the Fano resonance picture.

\section{Summary}
We have considered the behavior of an Anderson localized fermionic chain perturbed by an additional central site. This model represents a simplified (non-interacting) fermionic version of central spin models. The coupling to the central site was chosen as $m/\sqrt{L}$, corresponding to a fixed tunneling rate in the thermodynamic limit. 

We found that irrespective of the coupling strength $m$, wave functions of the central site model keep the localized statistics. Nevertheless, the wave functions loose their local character, as the central site typically causes resonances between localized orbitals that are far away. Instead, these rare resonances are responsible for multifractal wavefunctions that consist of many fractal replicas of the original Anderson localization peak.

The statistics of level spacings and the dynamics of the model revealed two different regimes. In the weak coupling regime, $m\ll W$ (with the disorder strength $W$), the central site fails to introduce the level repulsion and most of the eigenvalues of the system are Poisson-distributed. This regime shows logarithmic in time growth of the entanglement entropy in a quench protocol where one starts with an initially unentangled state. Since this a non-interacting model, the entanglement growth is caused by the slow ``leaking'' of particles between distant resonant sites mediated by the central site. 

The second regime, when the coupling to the central site is strong, $m\geq W$, is characterized by level repulsion in major parts of the spectrum. In this case the model can be mapped to the Fano resonance problem. The dynamics is then  characterized by a ballistic motion of entanglement.

Overall our findings support the intuition that a single degree of freedom, even if it is coupled non-locally to the Anderson insulator, is not sufficient to delocalize the system. At the same time, this non-local nature of the coupling allows for transport between initially localized orbitals via the central site. Surprisingly, dynamical probes uncover different transport regimes which are intimately connected to the presence/absence of level repulsion in the system. In particular, the logarithmic entanglement spreading which is usually attributed to the many-body localization, emerges in the present model as a signature of almost resonant sites. 

In a broader perspective, the present model illustrates emergence of distinct dynamical regimes from a perturbation of a localized system. These new dynamical regimes arise due to the absence of level repulsion, which is a fingerprint of localized systems.  It is an interesting and open question to extend the present analysis to the interacting regime. The natural interacting generalization of a non-interacting central site model is the central spin model which has a plethora of physical realizations.  We expect that studies of an interacting model would be useful for understanding and better control of spin relaxation mechanisms in quantum dots and could provide an alternative platform for studies of dynamics of interacting disordered systems.



\section{Acknowledgements} We would like to thank Dmitry Abanin, Christophe De Beule, Joel Moore,  Romain Vasseur, and Norman Yao for many stimulating discussions. Financial support has been provided by the Deutsche Forschungsgemeinschaft (DFG) via grant TR950/8-1, SFB  1170  ``ToCoTronics'' and the ENB Graduate School on Topological Insulators. M.S. was supported by Gordon and Betty Moore Foundation's EPiQS Initiative through Grant GBMF4307. F.P. acknowledges support from the DFG Research Unit FOR 1807 through grants no. PO 1370/2-1.

\appendix

\section{Mapping to Fano resonance problem}
\label{app:fano}
\subsection{Spectral weight and position of the resonance}
\label{app:spectral_weight}
Physically, the CSM resembles the physics of  the Fano resonance, when we have a single site coupled to a continuum. Then, there is a leading order self-energy correction, which determines the new position of the resonance,
\begin{equation}\label{Eq:Sigma}
  \Sigma(i\omega) 
  =
 \frac{m^2}{L} \sum_\alpha \frac{1}{i\omega -\varepsilon_\alpha},
\end{equation}
where $\varepsilon_\alpha$ labels energies of localized eigenstates. Approximating the sum with an integral, we deduce the following self-consistent equation for the new position of the resonance $\omega_r$ expressed via the dimensionless variable $x$ as $\omega_r = W x$: 
\begin{equation}\label{Eq:omegar2}
x
=
\frac{m^2}{2W^2} \ln \frac{x+1}{x-1}
\end{equation}
This formula holds for the two symmetric solutions $x_1 = -x_2$ outside the band, i.e. $\abs{x_i}>1$. In the following, we study the positive solution $x>1$. 
If the coupling is strong, i.e. $\mu:=\abs{m/W} \gg 1$, we find that $x\gg1$. Then, we expand the log and get
\begin{equation}\label{Eq:x-large}
 x  \approx \mu + \frac{1}{6 \mu} \gg 1.
\end{equation}
On the other hand, if the coupling is weak, $\mu\ll 1$, we obtain $x-1\ll 1$, so we have a resonance which is very close to the band edge,
\begin{equation}
 x-1 = 2 e^{-2/\mu^2 } \ll 1.
\end{equation}
Depending on $\mu$, we therefore have a very different spectral weight on the resonance, determined by
\begin{equation}\label{Eq:Z-def}
 Z = \left \vert 1 - \partial_\omega \text{Re}\,\Sigma(\omega)\vert_{\omega=\omega_r} \right\vert^{-1} 
 =\left \vert 1 + \frac{\mu^2}{x^2-1} \right\vert^{-1}.
\end{equation}
For  strong coupling $\mu\gg 1$, when $x\gg1$, we obtain:
\begin{equation}\label{Eq:Z-large}
Z \approx \frac{1}{2} - \frac{1}{6\mu^2},
\end{equation}
so that the spectral weight is almost $1/2$. The spectral weights of both symmetric solutions thus add up to $Z_\text{tot} = 2Z= 1-1/(3\mu^2)$, such that the spectral weight that remains within the continuum $[-W,W]$, i.e. $1-Z_\text{tot}$, is very small and scales as $1/\mu^2$.  On the other hand, for $\mu\ll1$, we have 
\begin{equation}\label{Eq:Z-small}
 Z  \approx \frac{4}{\mu^2}e^{-2/\mu^2}\ll 1,
\end{equation}
so that the spectral weight that remains in the continuum is close to one. However, in both cases, we recover that the spectral weight remaining in the bath does not scale with the system size. 

Since the only way to get delocalized states arises from the hopping via the central site, we conclude that the constant weight, when distributed among $L$ states in the continuum, will give a $\propto 1/L$ constant background of the wave function. The same result will be derived below in a complementary way from a three-site model.

\subsection{Dynamics and breakdown of the Fano description}
\label{app:fano_breakdown}
Using self-consistent perturbation theory, we can calculate the expansion of eigenstates with the central site over unperturbed eigenstates. This expansion reads: 
\begin{equation}\label{Eq:expansion}
 c_0 = \sum_i \nu_i  \alpha_i,
 \qquad
 c_i = \sum_j W_{ij}  \alpha_j.
\end{equation}
The coefficients $W_{ij}$ are expressed via $Z_i=Z(\epsilon_i)$ as 
\begin{eqnarray}\label{Eq:Zeta}
 W^2_{ij} &=& \frac{A^2 \nu^2_j}{(\epsilon_i-\epsilon_j)^2} + \delta_{ij} \frac{Z_i^2}{Z_i^2+(2\pi/\delta)^2},
\end{eqnarray}
where $A=m/\sqrt{L}$ in our case. 
Note, that the self-energy in Eq.~(\ref{Eq:Z-def}) defining $Z$ is given by:
\begin{equation}\label{Eq:SigmaB}
\Sigma(\omega) 
=
\frac{A^2}{\delta} \mathop{\mathrm{arctanh}} \frac{\omega}{W} \quad \text{for}\quad |\omega|< W.
\end{equation}
Now, we are interested in the amplitude of the wave function, that is not located on the initial eigenstate with energy $\epsilon_i$, which we denote as $n_\epsilon$ (index $i$ is omitted for brevity). To calculate this amplitude, we evaluate 
\begin{multline}\label{Eq:n-i}
 n_\epsilon= \sum_{j\neq i} |W_{ij} |^2
 =
 \sum_{j\neq i}   \frac{1}{(\epsilon_i-\epsilon_j)^2} \frac1{Z_j^2 + (2\pi/\delta)^2}
\\ =
 \frac{1}{\delta}\int d \omega    \frac{1}{(\omega-\epsilon)^2} \frac{m^4/L^2}{[\omega-\Sigma(\omega)]^2 + (\pi m^2/W)^2}
 .  
\end{multline}
To deduce the dependence on time, we can limit the sum in Eq.~(\ref{Eq:n-i}) to states $j$ with energy difference such that $|\epsilon-\epsilon_j|\geq 1/t$:
\begin{multline}\label{Eq:n-i-t}
 n_\epsilon(t)= \sum_{j\neq i} |W_{ij} |^2
 \\
 =
 \frac{2}{\delta}\int^\infty_{1/t} d \omega    \frac{1}{\omega^2} \frac{m^4/L^2}{[\omega-\epsilon-\Sigma(\omega-\epsilon)]^2 + (\pi m^2/W)^2}
 .  
\end{multline}
The integral yields the asymptotic expression
\begin{equation}\label{Eq:n3}
n_\epsilon(t) = \frac{1}{3} \frac{m^4}{L W}\ t^3
\end{equation}
at small times $t\ll 1/\epsilon_i$.  For longer times, when $W/m^2> t > 1/\epsilon$, assuming that $\epsilon>m^2/W=\epsilon_*$, we get: 
\begin{equation}\label{Eq:n-int-times}
n_\epsilon(t)\propto \frac{m^4}{W L}  \frac{t}{\epsilon^2},
\end{equation}
the linear growth of particle density that is located away from the initial eigenstate. Note, that for $t=1/\delta$ this expression gives us the saturation value that can be directly obtained from Eq.~(\ref{Eq:n-i}):
\begin{equation}\label{Eq:n-inf-time}
\bar n_\epsilon= n_\epsilon(1/\delta)=
 \frac{m^4}{4 W^2} \frac{2}{\epsilon_i^2+ (\pi m^2/W)^2}\propto  \frac{m^4}{W^2} \frac{1}{\epsilon_i^2}.
\end{equation}
We see that  $\bar n_\epsilon$ is suppressed as $1/\epsilon^2$ when the absolute value of energy is bigger than the crossover scale, $|\epsilon|>m^2/W =\epsilon_*$ defined in the main text. Such dependence would lead to a ballistic spreading of particles, and consequently a linear spreading of  entanglement entropy. 

In all above considerations, we however ignored the almost resonant pairs of sites. This is a legitimate assumption when the initial bath of states has level repulsion, so that the probability to have degenerate energies is vanishing. It is this assumption that breaks down for the present model, specifically outside the energy window~$[-\epsilon_*,\epsilon_*]$. In order to rigorously consider the physics emerging from such resonances,  we present a toy model which allows for an analytical understanding thereof in the next section.
\section{Beyond Fano picture: three-site model}
\label{app:three_site}
\subsection{Time dependent perturbation theory}
\label{app:perturbation_theory}
In order to derive the logarithmic growth of entanglement entropy, we study the
motion of a single fermion below. In particular, we calculate $n_\epsilon(t)$ for small times perturbatively and compare our analytical results with numerical data for the central site model.

We have found out that it is important to treat the coupling between initial
site and central site exactly. This is because the two hybridized states, a mixture of the initial site and the central site, contain the essential physics how the excitation moves into the remainder of the system. Thus, we consider the following, unperturbed Hamiltonian
\begin{equation}H_0 = \epsilon c_1^\dagger c_1 + A c_1^\dagger c_0 + A c_0^\dagger c_1 + \sum_{i\geq 2} h_i c_i^\dagger c_i \; ,
\end{equation}
where $A$ is the coupling to the central site. In the main text, we use $A=m/\sqrt{L}$.
If we diagonalize this Hamiltonian, we obtain
\begin{equation}H_0 = \lambda_+ f_0^\dagger f_0 + \lambda_- f_1^\dagger f_1 + \sum_{i\geq 2} h_i c_i^\dagger c_i
\end{equation}
with $\lambda_{\pm} = \frac{\epsilon}{2} \pm \sqrt{A^2 + \frac{\epsilon^2}{4}}$. Within this basis, the (perturbatively treated) coupling term becomes
\begin{align}
V &:= \sum_{i\geq 2} (c_i^\dagger c_0 + c_0^\dagger c_i)\\
& =n_- \sum_{i\geq 2} (f_i^\dagger f_1 + f_1^\dagger f_i) + n_+\sum_{i\geq 2} (f_i^\dagger f_0 + f_0^\dagger f_i)
\end{align}
and the initial state can be written as
\begin{equation}
\ket{\psi(t=0)} = c_1^\dagger \gs = \frac{1}{A}(n_+ \lambda_+ f_0^\dagger + n_- \lambda_- f_1^\dagger) \gs
\label{eq:initialstate}
\end{equation}
with $n_\pm = 1/\sqrt{1+\lambda_\pm^2/A^2}$ and $\gs$ is the (empty) vacuum state.
The total Hamiltonian is then given by $H=H_0 + A V$
and we are interested in the probability $\abs{\braket{n}{\psi(t)}}^2$ to
find the Fermion on another ring site $n\geq 2$ with potential
$h_n\in [-W,W]$. To solve this problem, we use the Ansatz
\begin{equation}
\ket{\psi(t)} = \sum_n  b_n(t) \exp(-ih_n t) \ket{\phi_n},
\end{equation}
where $\ket{\phi_n} = c_n\gs$ for $n\geq 2$ and $\ket{\phi_n} = f_n \gs$ for $n\in\{0,1\}$ are the eigenstates of $H_0$. Inserting $\ket{\psi(t=0)}$ in the Schr\"odinger equation,  this yields
\begin{equation}
i \frac{\tot{}}{\tot{t}} b_n(t) = \sum_k e^{i (h_n - h_k) t} V_{nk} b_k(t),
\end{equation}
where $h_i = \lambda_{\pm}$ for i $\in\{0,1\}$ and $V_{nk} = \bra{\phi_n} V \ket{\phi_k}$. This exact solution is now approximated by the expansion (in powers of $A$)
\begin{equation}
b_n(t) = b_n^{(0)}(t) + Ab_n^{(1)}(t) + A^2 b_n^{(2)}(t) + \ldots,
\end{equation}
for which we find 
\begin{align}
i \frac{\tot{}}{\tot{t}} b^{(r)}_n(t) &= \sum_k e^{i (h_n- h_k) t} V_{nk} b_k^{(r-1)} \quad r \geq 1\\
i \frac{\tot{}}{\tot{t}} b^{(0)}_n(t) &= 0.
\end{align}

\begin{figure}
\centering
\includegraphics[width = 0.98 \linewidth]{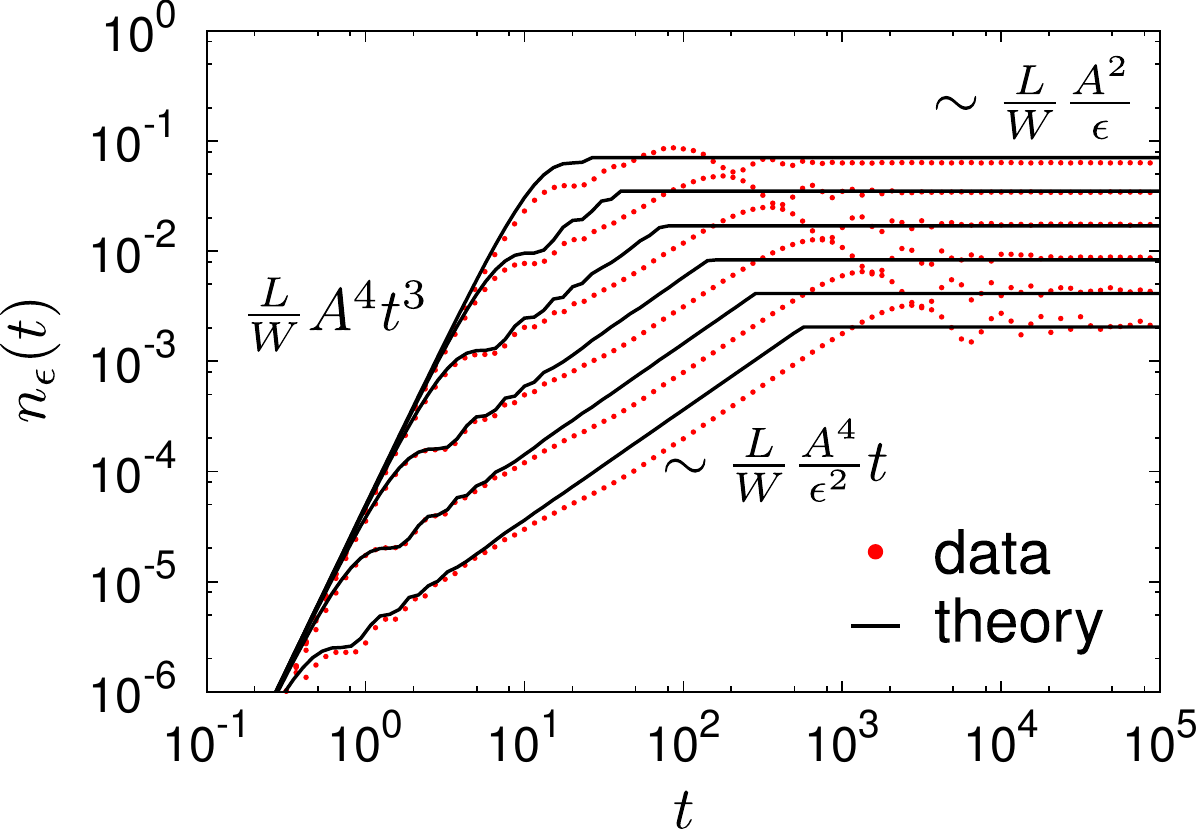}
\caption{The probability $n_\epsilon(t)$, that a particle, initially placed at a site of potential $\epsilon$, left its site and is located at another ring site has different behavior depending on the value of $\epsilon$. Upon increasing $\epsilon \in \{0.4, 0.8,1.6,3.2, 6.4 \}$ from top to bottom, we see that the time interval where $n_\epsilon(t)$ has a linear in time growth increases, while the saturation value decreases.  
}
\label{fig:single_fermion}
\end{figure}

Using the initial conditions (see Eq.~(\ref{eq:initialstate})) $b_0^{(0)}(t=0) = n_+ \lambda_+ / A$, $b_1^{(0)}(t=0) = n_- \lambda_- / A$, and $b_n^{(0)}(t=0) = 0$ for all $n\geq 2$, this implies
\begin{equation}
i \frac{\tot{}}{\tot{t}} b^{(1)}_n(t) = \frac{1}{A}\left(\lambda_+ n_+^2 e^{i(h_n - \lambda_+)t} + \lambda_- n_-^2 e^{i(h_n - \lambda_-)t}\right)
\end{equation}
and
\begin{equation}
i \frac{\tot{}}{\tot{t}} b^{(2)}_n(t) = \sum_{k\in\{0,1\}} e^{i (h_n - h_k)t} V_{nk} \underbrace{b^{(1)}_k(t)}_{=0} = 0.
\end{equation}
The time dependent probability that the Fermion
is present at site $n\geq 2$ is then well aproximated  by
\begin{widetext}
\begin{align}\nonumber
n_\epsilon(t,h_n) := &\abs{\braket{\phi_n}{\psi(t)}}^2\approx \abs{b_n^{(0)} + A b_n^{(1)}(t) + A^2 b_n^{(2)}}^2 \\
=& \frac{2n_+^4 \lambda_+^2 \left[ 1 - \cos(h_n - \lambda_+ t)\right]}{(h_n - \lambda_+)^2} + \frac{2n_-^4 \lambda_-^2 \left[ 1 - \cos(h_n - \lambda_- t)\right]}{(h_n - \lambda_-)^2}\\ 
& + \frac{2 n_-^2 n_+^2 \lambda_- \lambda_+ \{1 - \cos[(h_n - \lambda_+)t]- \cos[(h_n - \lambda_m)t]  + \cos[(\lambda_m - \lambda_p)t]\}}{(h_n - \lambda_+)(h_n - \lambda_-)}\nonumber
\end{align}
\end{widetext}
With this result at hand, we sum the contributions of all $L$
sites with random potential $h_n$. In the numerical simulation,
this is automatically done by considering many independent particles
at a time and by averaging over disorder.
Mathematically, we do this by
\begin{equation}
n_\epsilon(t) = \frac{L}{2W} \int_{-W}^{W} \tot{h} n_\epsilon(t,h).
\end{equation}
Using the principal value of the integral and extending the integration boundaries to  $(-\infty,\infty)$, which corresponds to a small mistake for $A\ll W$, the integral becomes 
\begin{equation}
n_\epsilon(t) \approx \frac{L}{2W} \frac{4 A^4 \pi}{\kk^3} \left(t \kk  - \sin(t \kk )\right),
\end{equation}
where $\kk=2\sqrt{A^2 + \frac{\epsilon^2}{4}}$ is the level splitting. For times $t \gg \frac{1}{\kk}$, we hence have derived the linear growth of $n_\epsilon(t)$. For times $t\ll \frac{1}{\kk}$, we find an initial cubic growth of $n_\epsilon(t)$, which coincides with Eq.~(\ref{Eq:n3}), but is not important for the logarithmic growth of entanglement entropy. In Fig.~\ref{fig:single_fermion}, we compare the perturbation theory with exact numerics and find good agreement.

\subsection{Saturation values}
\label{app:saturation_values}
In this section, we specify the limit\hfill
\begin{equation}\bar{n}_\epsilon :=\lim_{T\to\infty}\frac{1}{T}\int_{0}^T\tot{t} n_\epsilon(t).
\end{equation}
Recall that $n_\epsilon(t)=\sum_{i\geq2}^N \bra{\psi(t)}c_i^\dagger c_i\ket{\psi(t)}$ is the probability that the single fermion left its initial site and is now present in the bath.
In order to derive $\bar{n}_\epsilon(t)$, we first resolve it by energy, i.e.
\begin{equation}
\bar{n}_\epsilon (E) \tot{E} = \sum_{\substack{i\geq2\\  \mathclap{E \leq h_i < E+\tot{E}} }}^N  \lim_{T\to\infty}\frac{1}{T}\int_{0}^T\tot{t} \bra{\psi(t)}c_i^\dagger c_i\ket{\psi(t)}.
\end{equation}
As in the previous chapter, we treat the initial site $i=1$ and the central site $i=0$ exactly, i.~e. we diagonalize their two-site Hamiltonian $H_0 = 0 c_0^\dagger c_0 + \epsilon c_1^\dagger c_1 + A (c_1^\dagger c_0 + c_0^\dagger c_1)$ and find the two hybridized states $f_0$ and $f_1$. Subsequently, we couple these two states independently to a single bath state of energy $E$.

In a simple two site problem with energy gap $\Delta_{if}$ and coupling constant $A$, the probability to find an excitation outside its initial position is (averaged over time) given by
\begin{equation}
\overline{P}_{if} = \frac{2A^2}{\Delta^2_{if}+ 4A^2 }.
\end{equation}
The probability we seek is thus well approximated by
\begin{multline}
\bar{n}_\epsilon(h_i) = \abs{\braket{f_0}{\psi_0}}^2 \overline{P}_{f_0 i} + \abs{\braket{f_1}{\psi_0}}^2 \overline{P}_{f_1 i}\\
= \frac{\lambda_+^2 n_+^2}{ A^2} \,\frac{2 n_+^2 A^2}{(\lambda_+ - h_i)^2  + 4A^2 n_+^2} \\
+ \frac{\lambda_-^2 n_-^2}{ A^2} \,\frac{2  n_-^2 A^2}{(\lambda_- - h_i)^2  + 4A^2 n_-^2} \;.
\end{multline}
This probability is derived for one single bath state. We actually have $L-1$ bath states uniformly distributed in the energy window $[-W,W]$, resulting in a level density of $\frac{L-1}{2W}$. The average effect of all $L-1$ bath sites is thus given by
\begin{multline}
\bar{n}_\epsilon = \int_{-\infty}^\infty \tot{h} \frac{L-1}{2W} \bar{n}_\epsilon(h) \label{eq:sat_values}
= \frac{L-1}{2W} \frac{\pi}{A} \left[ \lambda_+^2n_+^3 + \lambda_-^2 n_-^3\right]\\
\overset{A\ll\epsilon}{=} \frac{L-1}{2W} \frac{A^2 \pi}{\epsilon} + \mathcal{O}(A^3) 
\end{multline}
Here, we are weakly overestimating the probability $\bar{n}_\epsilon$ due to two reasons. First, the integral should be taken from $-W$ to $W$. However, as we explicitly checked, the error caused by extension of integration limits to infinity is negligible for  small $A\ll \epsilon <W$. Secondly, although the assumption that all bath states are uncoupled is valid for small $A$, it underestimates the physical outcome. In the numerical simulation, the particle has more than one possibility to enter the ring sites. The ``fraction'' of the particle that goes to an $E_1$ ring site can no longer go to an $E_2$ ring site and vice versa. Thus, the analytical theory, which assumes only one bath state at a time, is overestimating the probability $\bar{n}_\epsilon$.
In Fig.~\ref{fig:single_fermion}, we compare the derived saturation value with the numerical gained data and find perfect agreement in the regime $A\ll \epsilon$ (lower curves).

\begin{figure}
\center
\includegraphics[width=0.98\linewidth]{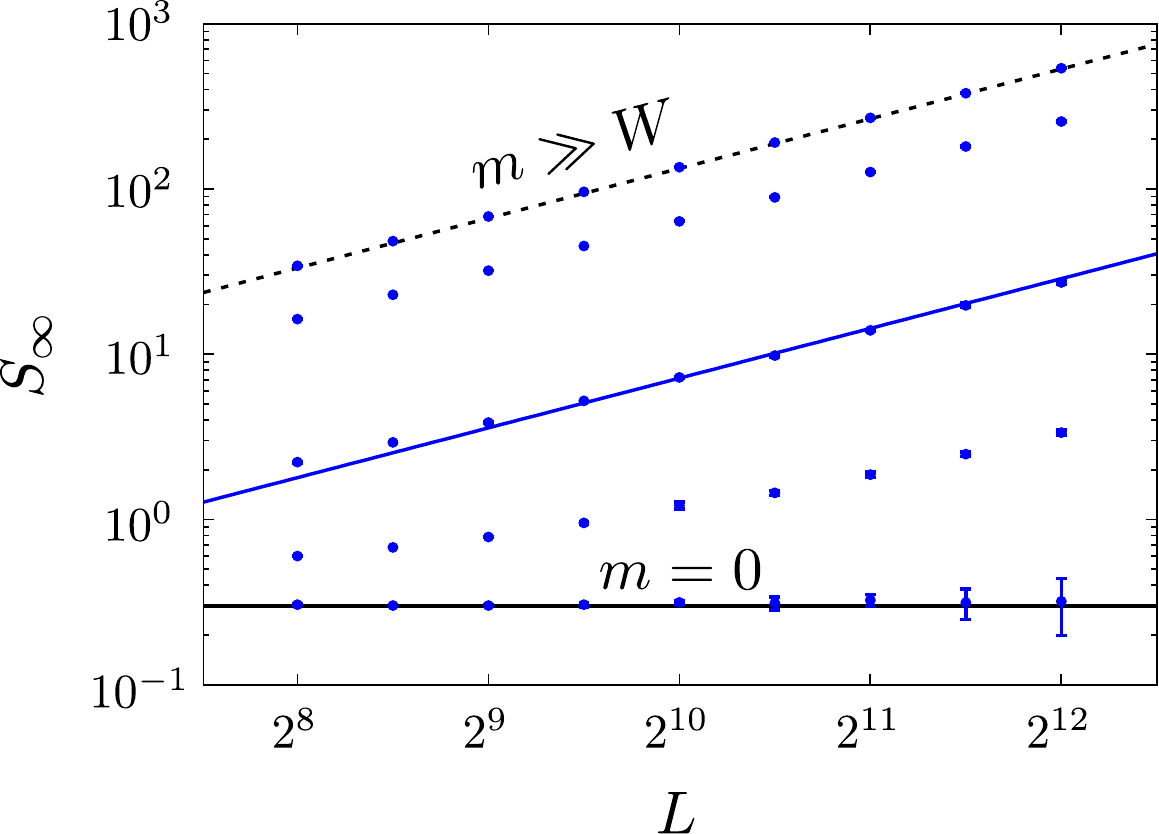}
\caption{Saturation value of the entanglement entropy $S_\infty=\lim_{t\to\infty}S(t)$ for different system sizes $L$ and coupling values $m\in\{0,0.25,1,5,20\}$ (bottom to top) and $W=10$. For all coupling constants $m>0$ we find a linear scaling $S_\infty \sim L$, e.g. we compare the data set of $m=1$ with $S_\infty =0.007L$ (blue line). This feature, together with the logarithmic in time growth of $S(t)$ (see Fig.~4) has never been observed before in a noninteracting system.}
\label{fig:entsaturation}
\end{figure}

Eq.~(\ref{eq:sat_values}) is not only completing the derivation of the logarithmic motion, but it also shows that saturation values should grow linear with system size. Indeed, we find that also the entanglement entropy saturates at values that grow linear with system size, see Fig.~\ref{fig:entsaturation}.

\subsection{Constant background level}
\label{app:constant_background}
Previously, we have described that all eigenstates $\ket{E_l}$ of the full Hamiltonian are localized at a site $l$. On sites $k$ far away from $l$, we see on average a constant probability $\abs{\braket{k}{E_l}}^2$ that the excitation $\ket{E_l}$ is measured at site $k$. With the above presented toy model, it is possible to deduce the correct scaling behavior of $\abs{\braket{k}{E_l}}^2 \sim \frac{m^2}{L W^2}$, which will be outlined next.

Let us consider again a single particle, initially placed at site $k$ with potential $\epsilon$. The time-averaged density matrix
\begin{equation}
\omega = \lim_{T\to\infty}\frac{1}{T}\int_0^T \tot{t}\rho(t) = \sum_{l} (\rho_0^E)_{ll} \ket{E_l}\bra{E_l}
\end{equation}
with $\bar{n}_\epsilon(E_l)\equiv (\rho_0^E)_{ll}:=\abs{\braket{E_l}{\psi(t=0)}}^2=\abs{\braket{E_l}{k}}^2$
contains the probability $\bar{n}_\epsilon(E_l)$ to measure the initial excitation at long times with an energy $E_l$, which is the overlap $\abs{\braket{k}{E_l}}^2$ we search for. Before, we have been interested in the total probability that the initial excitation is somewhere in the bath and found
\begin{equation}
\bar{n}_\epsilon = \int_{-\infty}^\infty \tot{h} \frac{L-1}{2W} \bar{n}_\epsilon(h).
\end{equation}
Now, we want the \emph{average} value of $\abs{\braket{E_l}{k}}^2= \mean{\bar{n}_\epsilon(E)}_E$, which is for $L-1$ bath states consequently given by
\begin{equation}
\mean{\abs{\braket{E_l}{k}}^2}_l = \frac{1}{L-1}\int_{-\infty}^\infty \tot{h} \frac{L-1}{2W} \bar{n}_\epsilon(h) = \frac{A^2\pi}{2W \epsilon}
\end{equation}
Finally, we have to average over the potential $\epsilon$ of the initial site $k$, which gives
\begin{align}
\mean{\abs{\braket{E_l}{k}}^2}_{l,k} &\approx 2 \frac{1}{2W}\int_{\alpha>0}^W  \tot{\epsilon}\frac{A^2\pi}{2W \epsilon} = \frac{A^2\pi}{2W^2}\ln\frac{W}{\alpha}.
\end{align}
In the main text, we used $A = \frac{m}{\sqrt{L}}$ and found the constant background level $\sim\frac{m^2}{L W^2}$, which coincides with this result, as the logarithmic contribution of the disorder $W$ is weak compared to the quadratic dependency and cannot be distinguished numerically.

\section{Relation between entanglement entropy and particle transport}
\label{app:transport_to_entanglement}
The entanglement entropy between two bipartitions $A$ and $B$
of a Hilbert space $\mathcal{H}$ is usually measured by
means of the von Neumann entropy of the reduced
density matrix $\rho_A=\text{tr}_B[\rho(t)]$, i.e.
\begin{equation}
S_A(t) = -\tr{\rho_A \ln \rho_A},
\end{equation}
where $\rho(t)$ describes the state of the system.
For pure states $\rho=\ket{\psi(t)}\bra{\psi(t)}$, $S_A(t)=S_B(t)$ for all $t$
and bipartitions $A,B$, which can easily be shown using the
Schmidt decomposition $\ket{\psi}= \sum_i \sqrt{\lambda_i} \ket{i}_A \otimes \ket{i}_B$.

For lattice systems of $L$ sites and free particles,
it is useful to describe a quantum state not with a
vector $\ket{\psi}\in\mathcal{H} = \mathbb{C}^{2L}$,
but by the correlation matrix
\begin{equation}
C_{ij}(t) = \bra{\psi(t)} c_i^\dagger c_j \ket{\psi(t)}
\end{equation}
where $c_i$ is a annihilation operator on
site $i$, and thus, $C$ is a correlation
matrix of size $L\times L$  only.
The matrix $C$ contains all information
about the full state $\ket{\psi}$ or
$\rho$, because according to Wick's
theorem any correlation function splits
for free fermions into products
of two-point correlators $C_{ij}$. Hence,
also the entanglement entropy $S_A(t)$
is expressible by means of $C$, which is
\begin{equation}
S_A(t) = - \tr{ C_A \ln C_A + (\tone - C_A) \ln (\tone - C_A)},
\end{equation}
and $C_A$ is the part of $C_{ij}$
with $i,j\in A$. As the particles
are independent from each other,
it is sufficient to consider a
single particle. The entanglement
entropy is additive for independent
degrees of freedom, hence, the total
entanglement entropy sums up to
$S_A(t) = \sum_i S^i_A(t)$,
where $S^i_A(t)$ is the contribution
of the $i$th fermion. If only one fermion
exists (in a pure state), the matrix $C$
has only one nonzero eigenvalue, which is
equal to unity. Hence, $C$ is of rank 1 and all
possible submatrices $C_A$ are at maximum
of rank 1 as well. For any bipartition $A,B$,
the matrix $C_A$ has at maximum one nonzero
eigenvalue $\lambda$, which equals for the
same reason $\lambda = \tr{C_A}$.
The entanglement entropy thus simplifies to
\begin{equation}
S_A(t) = - \lambda \ln \lambda - (1-\lambda) \ln (1-\lambda).
\end{equation}
Using the relation $\lambda=\tr{C_A}$, we find
\begin{equation}
\lambda = \sum_{i\in A} \bra{\psi(t)} c_i^\dagger c_i \ket{\psi(t)} = \sum_{i\in A} \abs{\braket{i}{\psi}}^2 = n_\epsilon(t)
\end{equation}
where $n_\epsilon(t)$ is the
probability, that the single fermion
is present in the  subspace $A$.

In the paper, $n_\epsilon(t)$ is the
probability, that a single fermion,
initially placed at a site of potential
$\epsilon$ changed its bipartition at
time $t$. Hence, we do have access to
the entanglement entropy by
\begin{equation}
S_A(n_\epsilon(t)) = - n_\epsilon(t) \ln n_\epsilon(t) - (1-n_\epsilon(t)) \ln (1-n_\epsilon(t)).
\end{equation}
By this equation, we have identified a direct connection between $n_\epsilon(t)$ analytically derived in Eq.~(4) and the time evolution of the entanglement entropy $S_A(t)$. The $\epsilon$-averaged expression for $n_\epsilon(t)$ explains the logarithmic spread of a particle in the CSM, see Eq.~(5). Thus, the leading terms at intermediate time scales for the time evolution of $S_A(t)$ (which also shows a logarithmic time dependence) is directly related to this peculiar particle motion.

\end{document}